\documentclass[aps,prb,twocolumn,nowpacs,superscriptaddress,groupedaddress,amsmath,amssymb]{revtex4}
\usepackage {graphicx,epsfig,graphics,color}

\begin{document}


\title{Direct measurement of the magnetic penetration depth by magnetic force microscopy}

\author{Jeehoon Kim}
\email[Corresponding author: ]{jeehoon@lanl.gov} \affiliation{Los
Alamos National Laboratory, Los Alamos, NM 87545}
\author{L. Civale}
\affiliation{Los Alamos National Laboratory, Los Alamos, NM 87545}
\author{E. Nazaretski}
\affiliation{Brookhaven National Laboratory, Upton, NY 11973}
\author{N. Haberkorn}
\affiliation{Los Alamos National Laboratory, Los Alamos, NM 87545}
\author{F. Ronning}
\affiliation{Los Alamos National Laboratory, Los Alamos, NM 87545}
\author{A. S. Sefat}
\affiliation{Oak Ridge National Laboratory, Oak Ridge, Tennessee
37831}
\author{T. Tajima}
\affiliation{Los Alamos National Laboratory, Los Alamos, NM 87545}
\author{B. H. Moeckly}
\affiliation{Superconductor Technologies Inc., Santa Barbara, CA
93111}
\author{J. D. Thompson}
\affiliation{Los Alamos National Laboratory, Los Alamos, NM 87545}
\author{R. Movshovich}
\affiliation{Los Alamos National Laboratory, Los Alamos, NM 87545}

\date{\today}

\begin{abstract}

We present an experimental approach using magnetic force microscopy for measurements of the absolute
value of the magnetic penetration depth $\lambda$ in superconductors. $\lambda$ is obtained in a simple
and robust way without introducing any tip modeling procedure via direct comparison of the Meissner
response curves for a material of interest to those measured on a reference sample. Using a
well-characterized Nb film as a reference, we determine the absolute value of $\lambda$ in a
Ba(Fe$_{0.92}$Co$_{0.08}$)$_{2}$As$_{2}$ single crystal and a MgB$_2$ thin film through a comparative
experiment. Our apparatus features simultaneous loading of multiple samples, and allows straightforward
measurement of the absolute value of $\lambda$ in superconducting thin film or single crystal samples.
\end{abstract}

\maketitle

\section {Introduction}

The superconducting coherence length ($\xi$), magnetic
penetration depth ($\lambda$), and their anisotropy are
fundamental parameters that characterize superconducting
materials.\cite{Tinkham,Poole,Prozorov 2000,Bonalde 2000,Hardy
1993, blatter} A number of important properties, such as
superconducting critical fields and superconducting fluctuations
that affect vortex dynamics, can be obtained if the parameters
above are known.\cite{Tinkham,blatter} The value of $\xi$, which
depends on the Fermi velocity and the condensation energy of the
superconducting state, can be estimated from the upper critical
field ($H_{c2}$) using the Ginzburg-Landau theory. $\lambda$ is
related to the density of superconducting
electrons,\cite{Tinkham} and, in contrast to $\xi$, precise
determination of its absolute value is notoriously
difficult\cite{Tinkham,Poole} due to demagnetizing effects,
topography-related surface barrier, and inhomogeneity of the
sample. Typically $\lambda$ is calculated by indirect methods.
Several experimental techniques such as tunnel diode oscillator
(TDO),\cite{Prozorov 2000APL} temperature dependence of the flux
explusion in the Meissner state\cite{Civale 1987} or of the
reversible magnetization in the mixed state using superconducting
quantum interference device (SQUID) magnetometers,\cite{Thompson
1990} mutual inductance,\cite{Fiory 1988} surface
impedance,\cite{Mao 1995} infrared reflectivity,\cite{Basov 1995}
muon spin resonance ($\mu$SR),\cite{Sonier 1994,Sonier 2000}
nuclear magnetic resonance (NMR),\cite{Chen 2006} and magnetic
force microscopy (MFM)\cite{Roseman,Nazaretski 2009, Luan 2010,
Luan 2011} have been employed for measurements of $\lambda$ in
thin films and bulk samples. Each of these methods has its own
advantages as well as limitations, and some of them require
simulations with multiple fit parameters. For example, TDO
measurements depend on the quality and thickness of an Al film
deposited on top of a superconductor, which may not be fully
reproducible yielding errors in the obtained values of $\lambda$.
The temperature dependent Meissner response method is only
sensitive for thin films with magnetic filed parallel to the
surface. The reversible magnetization method works only in clean
samples or materials having an extended vortex liquid phase.
Mutual inductance and $\mu$SR techniques are limited to thin
films and bulk samples, respectively. $\mu$SR measures the second
moment of the magnetic field distribution around vortex,
consequently details of the vortex structure and the muon's
location affect experimental accuracy. Infrared reflectivity
allows measurement of the anisotropy $\lambda$ by polarization of
an incident light.

MFM has been widely used for studies of superconductors,
particularly for imaging and manipulation of vortices in
superconducting thin films and single crystals.\cite{Moser 1995,
Volodin 1998, Roseman 2002, Auslaender 2009, Straver 2008}
Recently, MFM was also used as a local probe of the magnetic
penetration depth,\cite{Nazaretski 2009, Luan 2010, Luan 2011}
where  the values of $\lambda$ were extracted by fitting either
an MFM signal from a single vortex\cite{Nazaretski 2009} (note,
this method doesn't work for systems with large values of the
magneic penetration depth \cite{Jeehoon PRB}) or a Meissner
response.\cite{Luan 2010, Luan 2011} These methods require a
thorough characterization of the probe tip and a well-defined
simulation model (dipole-monopole, monopole-monopole, $etc$.)
describing interactions between the MFM tip and a superconducting
sample. However, modeling procedures with multiple fitting
parameters introduce uncertainties in the resulting $\lambda$
values. In spite of technical difficulties MFM has certain
advantages {\it e.g.} localization of measured $\lambda$ values
providing a route to explore anisotropy of $\lambda$ throughout
the sample. In our approach, we obtain local values of $\lambda$
by directly comparing the Meissner curves for the sample under
investigation with those obtained for a reference sample with a
well-known $\lambda$; we emphasize that the measurements are done
using the same MFM tip during the same cool-down. We observed
strong dependence of the Meissner response curves on the shape of
the MFM tip. When the tip crashes or even slightly touches the
sample surface, Meissner response curves can change
significantly. During our experimental procedure we verify that
the MFM tip does not change its properties by comparing the
Meissner curve from a Nb reference sample before and after the
measurement. In this paper we demonstrate the validity of our
method by determining $\lambda$ in a
Ba(Fe$_{0.92}$Co$_{0.08}$)$_{2}$As$_{2}$ single crystal and a
MgB$_{2}$ film. Our results demonstrate that the same procedure
can be used in any single crystal or thin film superconducting
samples with a thickness greater than $\lambda$ (for thinner
samples $\lambda$ can be corrected in a straightforward
manner.\cite{Wei 1996, Xu 1995})

\section {Experiment}

\begin{figure}
\includegraphics [trim=0 0 0 16cm,clip=true,angle=0,width=8.5cm] {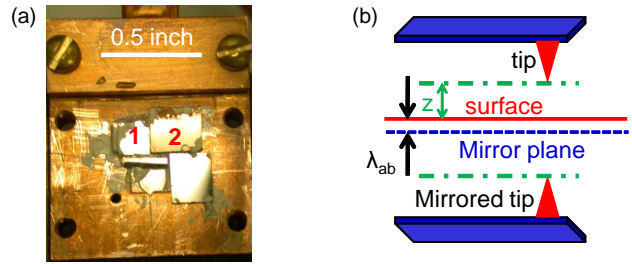}
\caption {\label{f:MFM} (Color online) (a) Sample holder with
multiple samples. The Nb thin film (300 nm) and the MgB$_2$ thin
film (500 nm) samples are labeled as 1 and 2, respectively. (b)
Schematic illustration of the Meissner experiment. The Meissner
response force between the probe tip and the sample can be
regarded as the force between the real tip and image tip at
$2(z+\lambda)$, when $z\gg\lambda$.\cite {Luan 2010}}
\end{figure}

The measurements described in this paper were performed in a
home-built low-temperature MFM apparatus.\cite{Nazaretski RSI
2009} We have developed an additional capability of mounting
multiple samples (including a reference sample), as shown in Fig.
\ref{f:MFM}(a), for acquiring a complete set of MFM data for each
of the samples within a single cool-down. The absolute value of
$\lambda$ is obtained by a simple comparison of the Meissner
curves for the Nb reference to those obtained in the sample of
interest. The Meissner response curve is first measured in a
homogeneous reference sample (Nb film) as a function of the
tip-sample separation. Then the cantilever is moved over the
sample of interest, and its Meissner response curve is obtained.
Direct comparison of these curves (comparative experiments)
yields the absolute value of $\lambda$ in a sample under
investigation. The value of $\lambda$ in the reference sample (Nb
film) was verified by a different MFM technique and a SQUID
magnetometry measurement.\cite{Nazaretski 2009} The reference Nb
thin film ($T_{c}\approx$ 8.8 K, where resistance drops to zero)
has a thickness of 300 nm and was grown by an electron beam
deposition. A Ba(Fe$_{0.92}$Co$_{0.08}$)$_{2}$As$_{2}$ single
crystal ($T_{c}\approx$ 22 K from specific heat capacity
measurements) was grown out of FeAs flux.\cite{Sefat,Kris} The
500-nm thick MgB$_2$ film ($T_c\approx$ 38.3 K, zero resistance
temperature) was grown by reactive evaporation.\cite{Moeckly
2006} The MFM measurements were performed using a high resolution
Nanosensors cantilever\cite {Nanosensors} that was polarized
along the tip axis in a 3 T field of a superconducting magnet.
The superconducting samples are zero field-cooled for the
Meissner experiment and are field-cooled in a field of few
Oersted, applied perpendicular to the film surface and parallel
to the probe tip, for imaging of vortices. All samples are
electrically grounded to eliminate a possible electric force
contribution from a stray charge on the sample to the magnetic
Meissner force.

\section {Results and Discussion}

\subsection {MFM measurement in the Nb film}

\begin{figure}
\includegraphics [trim=0 0 0 8cm,clip=true,angle=0,width=9cm] {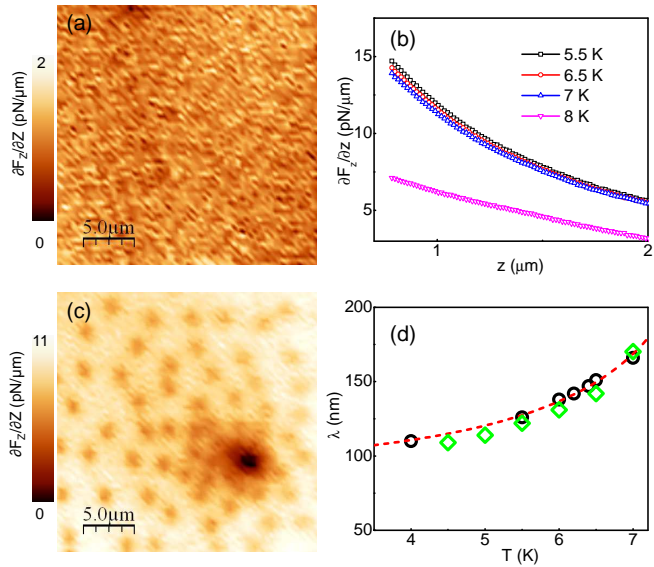}
\caption {\label{f:Nbresult}(Color online) (a) MFM image of the
Nb sample at $T$ = 4 K; no vortices are present in a $25~\mu
m\times 25~\mu m$ field of view. (b) Meissner response curves
taken over the scan area shown in (a). (c) The MFM image taken
with the same field of view as in (a) after the Meissner response
experiment at 8 K; vortices are nucleated by the magnetic field of
the MFM tip. (d) Temperature dependence of the penetration depth
$\lambda(T)$ in Nb obtained by overlaying Meissner curves on top
of the 4 K reference curve (black circles). Green squares
represent SQUID data; the red-dashed curve is a fit to the BCS
model.}
\end{figure}

As the magnetic tip approaches a superconducting sample, it
experiences a Meissner force (Meissner response) induced by the
shielding currents in the sample that screen the magnetic field
of the tip. The experimental procedure is as follows: First the
probe tip is brought close to the reference sample [Nb film,
position 1 in panel (a) of Fig.~\ref{f:MFM}] and the Meissner
response is recorded as a function of the tip-sample separation
at a certain temperature $T$. There should be no vortices present
in a large field of view of the sample, $i.e.$, the sample should
remain in a pure Meissner state. Figure~\ref{f:Nbresult}(a) shows
the Meissner state with no vortices present in the $25~\mu
m\times 25~\mu m$ field of view after a small compensation field
was applied above $T_c$ of Nb to compensate the remnant stray
field of the superconducting magnet.

Figure~\ref{f:Nbresult}(b) shows several Meissner response curves
for the Nb film reference sample between 5.5 K and 8 K. The
Meissner response force is a function of $\lambda$ and the
tip-sample separation $z$, $F_{M}=F(z+\lambda)$.\cite{Luan 2010,
Luan arxiv} For a given temperature $T$,
$F_{M}[z+\lambda(T)]=F_{M}[z+\lambda$(4~K)$+\delta\lambda]$, where
$\delta\lambda=\lambda(T)-\lambda$(4~K). Therefore, to determine
the $\lambda$ value at a particular temperature $T$,
$F_{M}[z+\lambda(T)]$ is shifted along the $z$ axis to coincide
with $F_{M}[z+\lambda$(4~K)], a reference curve measured at 4 K,
and the value of the shift yields the value of $\delta\lambda(T)$.

The Meissner response curve taken at 8 K shows a behavior very
different from the data taken at lower temperatures.
Figure~\ref{f:Nbresult}(c) shows the MFM image acquired at 4 K
after the Meissner measurement was performed at 8 K. The scan
areas in Fig.~\ref{f:Nbresult}(a) and Fig.~\ref{f:Nbresult}(c)
are the same. The vortices seen in Fig.~\ref{f:Nbresult}(c) were
generated by the field of the probe tip. They cause magnetic
field leakage and weaken the Meissner response (see
Fig.~\ref{f:Nbresult}(b), 8 K curve). Such a behavior is observed
when $\lambda$ is comparable to the film thickness, and the
magnetic field from the tip can not be fully screened.\cite{Wei
1996, Xu 1995} Special care must therefore be taken with the
Meissner technique in the vicinity of $T_c$ in thin films since
increasing $\lambda$ may take the sample out of the pure Meissner
state.

To verify the 4 K value of $\lambda$(4~K)$=110\pm$10 nm for the Nb
reference, measured previously using a different MFM
technique,\cite{Nazaretski 2009} we performed measurements of
$\lambda$ using the SQUID magnetometry. The Nb reference sample
used in MFM measurements (film L=3.2 mm $\times$ W=4.2 mm) was
oriented carefully with {\bf H} parallel to the surface along the
side L, using a home-built sample holder. By measuring the
transverse component $m_\bot$ of the magnetic moment {\bf m} in
the Meissner state (which should be zero in the case of perfect
alignment) we confirmed that the miss-orientation between {\bf H}
and the film surface was $\phi_{mis}\sim 0.2^{\circ}$. In this
configuration, in the Meissner state the component of {\bf m}
parallel to {\bf H} is $m_{\parallel}=(H/4\pi) \times LWd_{eff}$,
where the effective thickness
$d_{eff}(T)=d-2\lambda(T)\tanh(d/2\lambda(T))$ is smaller than
the geometrical thickness due to the field penetration from both
surfaces.\cite {Tinkham, Poole} We measured $m_\parallel$ versus
$H$ at several $T$, and from the slopes $dm_\parallel /dH$ we
extracted $\lambda(T)$. The main source of error in this method
is the spurious contribution to $m_\parallel$ due to the
projection of the Meissner signal arising from the transverse
field component, $\sim m_{\perp} \phi_{mis}$, which is $\sim$10
\% of $m_\parallel$, thus introducing an error of $\sim 5\%$.
However, it does not distort the functional dependence of
$\lambda(T)$. The absolute values of $\lambda(T)$ from these
SQUID measurements are marked with green squares in
Fig.~\ref{f:Nbresult}(d) and agree well with both the MFM data
(black circles) and the isotropic single-gap BCS model
(red-dashed curve).

\begin{figure}
\includegraphics [trim=0 0 0 12cm,clip=true,angle=0,width=9cm] {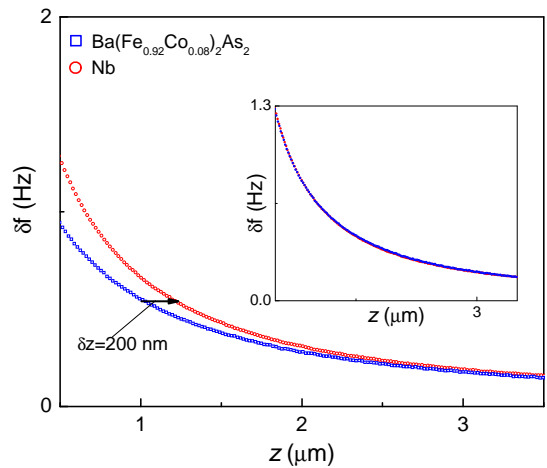}
\caption{\label{f:Ba122} (Color online) Meissner response curves
obtained from (a) the Nb reference, (b) a
Ba(Fe$_{0.92}$Co$_{0.08}$)$_{2}$As$_{2}$ single crystal at 4 K.
Different slopes of the Meissner curves obtained from each sample
indicate a systematic change of $\lambda$. Inset: The Meissner
curve of Ba(Fe$_{0.92}$Co$_{0.08}$)$_{2}$As$_{2}$ is shifted by
$z$=200 nm along the $z$ axis to overlay the Meissner curve of the
Nb reference sample. The difference of the penetration depths
$\Delta\lambda$= 200 nm can be obtained from the value of the
shift along the $z$ axis.}
\end{figure}

\subsection {Measurements of the absolute values of $\lambda$
in a Ba(Fe$_{0.92}$Co$_{0.08}$)$_{2}$As$_{2}$ single crystal}

In recent years iron-based pnictide superconductors have drawn a
great deal of attention since these systems exhibit
superconducting properties which are intermediate to conventional
BCS superconductors and high-T$_{c}$ cuprates. Magnetic
penetration depth of a Ba(Fe$_{1-x}$Co$_{x}$)$_{2}$As$_{2}$
system (BFCA), the so-called 122 family, was investigated using
variety of techniques including TDO,\cite{Gordon}
$\mu$SR,\cite{Ofer} and MFM.\cite{Luan 2011} The Stanford group
utilized an MFM technique and reported absolute values of
$\lambda$ in this system as a function of doping level. The
authors used fitting algorithms to approximate the tip
magnetization and calculate the values of $\lambda$.\cite{Luan
2011} We applied our direct (comparative) technique to
measurements of $\lambda$ in the very same system [
Ba(Fe$_{1-x}$Co$_{x}$)$_{2}$As$_{2}$ (x=0.08)] to demonstrate the
validity of our approach. Figure~\ref{f:Ba122} shows Meissner
curves as a function of the tip-sample separation $z$ obtained
from BFCA (blue squares) and a Nb reference (red circles),
respectively. The slow decay of a frequency shift in BFCA sample
compared to that in the Nb reference indicates greater values of
$\lambda$ in the BFCA sample. The expression for the Meissner
response force in Nb (assuming monopole-monopole interaction
between the tip and the sample) can be written as follows:
$F^{Nb}_{Meissner}=\frac{A\Phi_0}{(z+\lambda^{Nb}_{ab})^3}$,
where $A$ is a prefactor that reflects the sensor's geometry and
the magnetic moment, $\Phi_{0}$ is a single magnetic flux quantum,
$z$ is the tip-sample separation, and $\lambda^{Nb}_{ab}$ is the
in-plane magnetic penetration depth since the shielding current
runs within the basal plane. The Meissner force in BFCA has the
same functional form but different $\lambda^{BFCA}$:
$F^{BFCA}_{Meissner}=\frac{A\Phi_0}{(z+\lambda^{BFCA})^3}$. Two
Meissner curves become identical
($F^{BFCA}_{Meissner}=F^{Nb}_{Meissner}$) when the tip lift $z$
compensates for differences
$\lambda^{BFCA}-\lambda^{Nb}=\delta\lambda$. In other words, by
shifting the Meissner curve for BFCA along the $z$ axis to overlay
the Meissner curve measured for Nb, the $\delta\lambda$ value can
be extracted. By adding the shifted value $\delta z =
\delta\lambda = 200$ nm to the $\lambda^{Nb}$ (110 nm) one
obtains $\lambda^{BFCA}$=310$\pm$30 nm. This value is close to the
one reported previously.\cite{Luan 2011} The inset in
Fig.~\ref{f:Ba122} demonstrates that two Meissner curves overlay
each other very well after shifting the BFCA curve along the $z$
axis by 200 nm.

\begin{figure}
\includegraphics [trim=0 0 0 2cm,clip=true,angle=0,width=8.5cm] {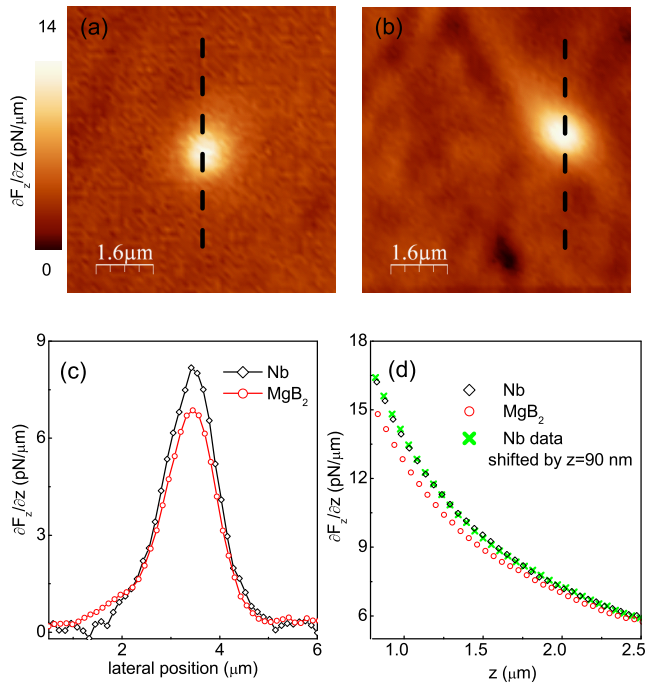}
\caption{\label{f:MgB2} (Color online) Single vortex images in
(a) the Nb thin film and (b) the MgB$_2$ thin film samples at $T$
= 4 K. Both images were taken with a tip-lift height of 400 nm.
(c) The single vortex profiles taken along the dotted lines in
(a) and (b). (d) Meissner response curves taken at 4 K in Nb
(black square) and MgB$_2$ (red circle) with the same experimental
condition. The green-crossed marks represent that the red circle
(MgB$_{2}$), after shifted by 90 nm along the $z$ axis, is
overlaid over the black square (Nb) to show the validity of our
approach.}
\end{figure}

\subsection {Measurement of the absolute value of $\lambda$ in a MgB$_{2}$ film}

We used the same approach to measure the absolute value of
$\lambda$ in a MgB$_{2}$ thin film sample. The Meissner response
curve taken at 4 K in the Nb was used as a reference Meissner
curve and compared to the Meissner response curve measured in
MgB$_{2}$. The offset between these two curves, shown in
Fig.~\ref{f:MgB2}(d), yields the absolute value of $\lambda$ at 4
K. The $\delta\lambda$ between the Nb and MgB$_2$ curves at 4 K
equals 90 nm yielding $\lambda$(4~K)$=200\pm$30 nm in MgB$_2$. Our
experimental error is $10\% - 15\%$, and depends on the magnitude
of $\lambda$ and the system noise level. Absolute values of
$\lambda$ in MgB$_{2}$ measured with various techniques range
from 40 nm to 210 nm.\cite{xxx1,Golubov, Dahm, Chen, Lee, Simon,
Finnemore} The large variation of $\lambda$ in MgB$_{2}$ may be
due to inclusion of impurities, such as C, N, and Al, which
replace either Mg or B and significantly affect the electronic
structure of the system due to the two-band nature of MgB$_{2}$.
The origin of the impurities-induced large value of $\lambda$ and
its temperature dependence obtained from the Meissner response
curves in the MgB$_{2}$ film taken at different temperatures (not
shown) will be described elsewhere.\cite{lambda-T}

We also imaged individual vortices in both Nb and MgB$_2$. Direct
comparison of the vortex profiles provides additional information
on the magnitude of $\lambda$. The maximum force gradient at the
center of a vortex, max($\partial f/\partial z$) (MFM is
sensitive to a force gradient), is proportional to
$(z+\lambda_{ab})^{-1/3}$ for a monopole-monopole model of the
tip-vortex interaction.\cite{Auslaender 2009,Straver
2008,Shapoval 2011} In this model, the larger value of $\lambda$
results in a smaller force gradient at the center of a vortex.
Figures~\ref{f:MgB2}(a)-(b) show well-isolated vortices acquired
in the Nb reference and the MgB$_{2}$ sample at 4 K. We estimate
the magnetic field to be no more than 0.2 Oe based on the field
calibration. Direct comparison of vortex profiles [see
Fig.~\ref{f:MgB2}(c)], taken along the dotted lines in
Figs.~\ref{f:MgB2}(a) and (b), shows that the force gradient in
the MgB$_{2}$ sample is smaller than that in the Nb sample. This
indicates a larger value of $\lambda$ in our MgB$_{2}$ film. It
is worth noting that the relatively high tip-sample separation
(400-nm tip-lift) and the tip geometry are responsible for the
broadening of the vortex force profiles in both Nb and MgB$_{2}$
samples in Fig.~\ref{f:MgB2}(c).

\section {Conclusion}

In conclusion, we have developed an experimental method and
apparatus to determine the absolute value of the magnetic
penetration depth $\lambda$ in superconducting samples by
comparing their Meissner response curves to those acquired for a
homogeneous Nb reference film. We used this method to obtain the
absolute value of $\lambda$(4 K)=$310\pm 30$ nm in a
Ba(Fe$_{0.92}$Co$_{0.08}$)$_{2}$As$_{2}$ single crystal,
consistent with the value reported by the Stanford group. We also
measured $\lambda$(4~K)$ = 200 \pm 30$ nm in a MgB$_2$ film. The
large $\lambda$ comes from the nature of the two band
superconductivity, and from inclusion of impurities such as C and
N. Our MFM apparatus allows us to simultaneously load and
investigate multiple samples (over ten samples can be studied at
once, providing an opportunity to explore the complete phase
diagram of a superconducting system), and most importantly to use
the same cantilever tip for both Nb reference and the samples
under investigation in a single cool-down. This capability
enables {\it in-situ} calibration of the MFM tip on a known
homogeneous Nb sample and does not introduce any additional
uncertainties due to modeling of the tip geometry and the
resulting tip field. The validity of our approach is established
by comparing the MFM and SQUID magnetometer measurements of the
temperature dependence of $\lambda$ in the Nb reference film. Our
experimental approach opens the possibility of measuring the
absolute value of $\lambda(T)$ in film and bulk superconducting
samples.

The authors thank Y. Q. Wang for the RBS measurement. Work at
LANL was supported by the US Department of Energy, Basic Energy
Sciences, Division of Materials Sciences and Engineering (MFM,
data analysis and manuscript preparation), and by T. Tajima 2010
DOE Early Career Award (SQUID measurements). Work at Brookhaven
(data analysis and manuscript preparation) was supported by the
US Department of Energy under Contract No. DE-AC02-98CH10886.
BaFe$_{2}$As$_{2}$ samples were grown at Oak Ridge National
Laboratory with the support of the Department of Energy, Basic
Energy Sciences, Materials Sciences and Engineering Division.
N.H. is member of CONICET (Argentina).

\end{document}